\documentclass[man]{apa6}
\usepackage{multirow}
\usepackage{xcolor}
\usepackage[utf8]{inputenc}
\usepackage{apacite}
\usepackage{graphicx}
\usepackage{subfigure}
\usepackage{rotating}

\usepackage{multicol}

\usepackage{multirow}

\usepackage{setspace}
\usepackage{adjustbox}

\leftheader{Participation and Privacy perception in virtual environments}

\shorttitle{Participation and Privacy perception in virtual environments}

\title{ {\bf Participation and Privacy perception in virtual environments: the role of sense of community, culture and sex between Italian and Turkish}}

\sixauthors{Andrea Guazzini}{Ayça Saraç}{Camillo Donati}{Annalisa Nardi}{Daniele Vilone}{Patrizia Meringolo}

\sixaffiliations{Department of Education and Psychology, and Center for the Study of Complex Dynamics (CSDC), University of Florence, Via di San Salvi 12, 50135, Firenze, Italy\\ andrea.guazzini@unifi.it}{Department of Educational Sciences, Çukurova University, Balcalı, Adana, Turkey.\\ asarac@cu.edu.tr}{Department of Education and Psychology, University of Florence, Via di San Salvi 12, 50135, Firenze, Italy\\ camillo.donati@unifi.it}{Department of Education and Psychology, University of Florence, Via di San Salvi 12, 50135, Firenze, Italy\\ annina.nardi@gmail.com\\ \ }{LABSS (Laboratory of Agent Based Social Simulation), Institute of Cognitive Science and Technology, National Research Council (CNR),  Via Palestro 32, 00185 Rome (Italy), and Grupo Interdisciplinar de Sistemas Complejos (GISC), Departamento de Matem\'aticas, Universidad Carlos III de Madrid, 28911 Legan\'es (Spain)\\ daniele.vilone@gmail.com} {Department of Education and Psychology, University of Florence, Via di San Salvi 12, 50135, Firenze, Italy\\ patrizia.meringolo@unifi.it}

\abstract{
Advancements in information and communication technologies have enhanced our possibilities to communicate worldwide, eliminating borders and making it possible to interact with people coming from other cultures like never happened before. Such powerful tools have brought us to reconsider our concept of privacy and social involvement in order to make them fit into this wider environment. It is possible to claim that the ICT revolution is changing our world and is having a core role as a mediating factor for social movements (e.g., Arab spring) and political decisions (e.g., Brexit), shaping the world in a faster and shared brand new way. It is then interesting to explore how the perception of this brand new environment (in terms of social engagement, privacy perception and sense of belonging to a community) differs even in similar cultures separated by recent historical reasons. Recent historical events may in effect have shaped a different psychological representation of Participation, Privacy and Sense of Community in ICT environments, determining a different perception of affordances and concerns of these complex behaviors.
The aim of this research is to examine the relation between the constructs of Sense of Community, Participation and Privacy compared with culture and sex, considering the changes that have occurred in the last few years with the introduction of the web environment. A questionnaire, including ad hoc created scales for Participation and Privacy, have been administered to $180$ participants from Turkey and Italy. In order to highlight the cultural differences in the perception of these two constructs, we have provided a semantic differential to both sub-samples showing interesting outcomes. The results are then discussed while taking into account the recent history of both countries in terms of the widespread of  new technologies, political actions and protest movements.

}


\begin{document}
\maketitle





\section*{Introduction}

The social nature of human beings has attracted the interest of researchers since the beginning of science. Being part of a social entity and actively participating in it is a part of everyone of us.  This natural tendency explained for the first time by Social Darwinism represents one of the environmental pressures that has probably affected most of the human evolution. Cooperation, competition and other social skills (e.g., social problem solving, social self- efficacy) as well as dynamics (e.g., individualism and pluralism, in-group favoritism) have determined human evolution and are still distinguishing our society more than other environmental features. Within such  complex dynamics, social and cultural groups have developed different strategies in order to maintain an efficient way to cooperate within their community while simultaneously adopting an effective strategy to compete with the other groups within their ecological niche.
  
Nowadays the revolution in information and communication technologies (ICT) is changing the old social equilibrium of the world by providing a brand new way to interact with others, new social structures and possibilities, as well as new "ecological niches" to be explored and exploited.

In the 1990s, Dunbar’s number theory set the cognitive limit of the number of people with whom one can maintain stable social relationships to $150$ \cite{dunbar1992neocortex, dunbar2009social}.
It is clear that with the massive use of social networks and online communication tools the number of social relations we deal with daily is increasing, which suggests the need to re-think the concepts of ecological niches as well as the concepts of community, participation and privacy.
Similarly, even the ecological system theory \cite{bronfenbrenner1992ecological} needs to be updated by considering the possible relations that can be developed through new scenarios such as the world wide web.
Such a brand new way to interact is in facts affecting our lives as well as the world organization, and presumably scientists should consider this as the most important environmental factor that will affect the human evolution in the future.
For example recent and important historical events (i.e. Arab Spring, Occupy Wall Street Movements, anti-coup movements in Turkey) took place thanks to the possibility provided by social networks to gather an extraordinary amount of people in such a brief amount of time: before the ICT revolution it would have been at least harder to make those movements happen. Sure enough social networks, and new media in general, are having a key role in driving debates and public opinion in consequence of socially relevant events and can be seen as mediators of the consequences of these.
At the same time recent historical events are shaping our way to interact with ICT tools so that similar cultures are approaching social networks and new media in a different way, depending on the degree of affordances and concerns related to their use.

In order to properly approach this environmental change, we need to consider the cultural differences characterizing the interaction between people within this brand new world (i.e., perception of privacy, virtual sense of community and web based participation) and distinguishing the new equilibria that interconnects human society through the world wide web \cite{prensky2001digital}.
Understanding the human factors involved in the use of social networks and online applications, such as mobile applications or crowdsourcing platforms, is now one the most challenging objectives for several disciplines \cite{arnaboldi2013egocentric, passarella2012ego}. In fact by considering the massive and daily use of web tools in our lives, we need  to understand how several well-known factors and socio-psychological dynamics in literature can be translated into virtual contexts. The opportunities in terms of communication and the sharing of knowledge that the "virtual world" is providing us are astonishing.  However in order to gain an actual comprehension of the usage of such tools, a deeper understanding of the complex dynamics behind those processes needs to be achieved. While a proper and comprehensive model for the human interaction with technologies is still under development, it can be useful to approach this research by using well known techniques adopted in the past in the "real world" as the cross-cultural analysis of features involved in the behaviours of interest.

Cross-cultural differences involved in the perception of behaviors, attitudes or emotions have frequently been a precious tool in order to understand peculiarities or commonalities in several areas of research . The enrichment provided by different cultures can, in fact, highlight peculiarities related to the context or the history of a population that may help researchers in modeling complex behaviors. Investigating the connections between well known constructs and taking into account the differences due to cultural specificity or recent historical events is necessary for researchers to understand new important features, excluding some others related to a specific context, but showing the important factors in common with the human race.

The main objective of our study is to measure the relationship between Sense of Community, Participation and Privacy compared to Culture and Sex, taking into account two culture (Italian and Turkish) similar for many features, but surely different for recent historical events. After a brief theoretical introduction to the constructs (Sense of Community, Participation and Privacy), we will describe our study in which 180 digital native participants (83 Italians and 97 Turkish People) have been asked to fill in an online questionnaire.

\subsection{Sense of Community}

Sense of community (SoC) is a construct largely used in community psychology, where it has been used to describe the complex relationship that an individual has while feeling as an active part of his community. Since the term “community” has a multi-level nature, which according to Brodsky and Marx (2002) can be referred to neighborhood, block groups, housing complexes, schools and cities, it has been studied in a large variety of contexts \cite{brodsky2002expanding, brodsky2011layers}. Given that several definitions have been taken into account, the variety of features that this construct can express for different individuals in different settings is phenomenal. Sarason defined it as the feeling that community members have about each other  \cite{sarason1974psychological, blanchard2008testing}, McMillan and Chavis (1986) claimed that SoC is a concept related to membership, emotional safety, identity, belonging and attachment to a group \cite{mcmillan1986sense}. The perception of similarity to others and willingness to communicate with people are also key features in defining SoC \cite{castellini2011sense}. According to these authors, when we feel part of a community we are more willing to share responsibility, to improve face to face relationships and to participate with others in social activities. On the other hand, several social scientists have shown that the increased complexity and change in technologies have affected the meaning and importance of communities \cite{brodsky2002expanding}. Relevant to this new approach to SoC, in which the contexts are evolving from a physical reality into a virtual scenario, all research is aimed at identifying a definition of the Sense of Virtual Community \cite{blanchard2002sense, rovai2005feelings, francescato2006evaluation}. The Sense of Virtual Communities has been found to be overlapping in the SoC in many features: for example participants in some studies by Francescato (2006; 2007) have shown similar levels in collaborative learning, developed self-efficacy, developed problem solving and perception of social capital when comparing face to face and online groups \cite{francescato2006evaluation, francescato2007developing}.

\subsection{Participation}

Due to its multi-disciplinary nature,  participation has been investigated through approaches coming from several academic disciplines (e.g. Sociology, Anthropology Psychology) even though a proper descriptive and comprehensive model is still lacking at the moment.  Given that, it can be analyzed starting from different levels: macro-social/institutional level, micro-social levels, psychological level. According to Cicognani and Zani (2011), when discussing participation on the macro-social/institutional level we refer to the electoral system, civic education, culture, religion, social-economic development and the history of the country in question \cite{cicognani2011civic}. On the micro-social level, contexts such as family, friends, school, and voluntary associations are taken into account . On the psychological level, Wilkenfeld et al. (2010) highlight the importance of participation, on both civic and political issues as well as on the cognitive and social development of adolescents \cite{wilkenfeld2010relation}. A key feature in promoting participation is the presence of a collaborative group sustaining the action. Sense of collective identity stimulates engagement in collective actions \cite{klandermans2002methods} as well as identification with the group \cite{cicognani2011civic}. To this effect, SoC \cite{mcmillan1986sense} has been proven to have a positive, bidirectional, association with participation \cite{simon1998collective}. One key feature related to participation is also the construct of empowerment \cite{rappaport1991healing} and as found in Zimmermann (2000) the action-oriented nature of empowerment finds a natural consequence in both individual and social participation \cite{zimmerman2000empowerment}.  Combining the contribution from  psychological literature and the recent findings coming from the renewed interest in this construct, especially from Information and Communication Technologies studies, several features have been taken into account to find the proper incentives promoting participation (inhibiting competition). For instance Gachter investigated the role of rewards and punishment, finding out that punishments can increase cooperation even in the long term, because the gains from cooperation itself give enough gains to became stable even if the punishment is removed \cite{gachter2008long}. In a more recent review on this topic, Gachter shows that even the size of the incentives has a crucial role in these dynamics. In a contest of rewards, a very large incentive can enhance a stronger cooperation, while with a higher punishment a low threat can be enough to reach the same result \cite {rand2009positive} \cite {sutter2010choosing} \cite {gachter2012social}

\subsection{Privacy}

With the increasing use of social networks (e.g., Facebook, Twitter, Google+), and collaborative web platforms (i.e., crowdsourcing tools) people  find the opportunity to share ideas and feelings with others on internet everyday . While exploring these brand new and innovative media, we have learned that we also need to take into account  the potential dangers that can occur while exchanging personal data online. Participating in social media is highly related to the feeling of safety that people experience while sharing their contents on the web. According to Song, Hao and Daqing (2013), people that have a higher trust in social networks are more likely to participate actively than those who have less \cite{song2013empirical}.  When people feel a risk about sharing personal data on social networks they do not contribute to them. As mentioned above in SoC, behavior in real physical contexts and web scenarios seem to be overlapping in some of its features. In fact, privacy is  a very important dimension of human life, affecting personal and social lives.  Since it involves the control of the amount of contact with others and the perceived safety of these interactions, Solove (2008) has found that when people do not meet their privacy needs, it may result in antisocial and stressful behaviors \cite{solove2008end}. 
Altman has defined privacy as “the selective control of access to information to the self or to one’s group”  \cite{altman1975environment}. According to Pedersen (1997) six types of privacy can be identified: solitude, isolation, anonymity, reserve, intimacy with friends and family \cite{pedersen1997psychological}. Privacy attitudes have been measured in a sample of $210$ men and $165$ women from high schools and colleges in Turkey \cite{rustemli1993privacy}. Results show that women have higher mean scores for measures of intimacy with friends and lower mean scores for isolation and reserve than men. There were no mean differences for solitude, intimacy with family, and anonymity.  Cross-cultural differences in the perception of privacy regulation have been investigated by Kaya and Weber in 2003, involving American and Turkish students \cite{kaya2003cross}. The result showed that American students desired more privacy in their residence hall rooms than Turkish students. Regardless of culture, males reported a greater desire for privacy than females. The relationship between privacy and security was investigated in a study regarding the educational use of cloud service such as social networks, Google drive and Dropbox \cite{arpaci2015effects}. Here, security was taken into account as the degree to which students believed that cloud-services were secure platforms for storing and sharing sensitive personal data. The results showed that the perception of a low level of security may have affected students’ attitudes towards using such services. In other words, students with low tolerance for technological risks may defer their use of these services and privacy concerns may then impede attitude towards educational use of cloud services. Several scales have also been recently developed in order to investigate privacy.
The Privacy Behavior Scale is a six-item, $5$ point Likert scale, assessing the need for privacy related to the use of Internet \cite{buchanan2007development}. The Privacy Attitude Scale \cite{buchanan2007development} measures privacy concerns for a number of Internet security topics, such as the use of e-mail with $16$ items on a $5$ point Likert scale. The Privacy Concerns Scale \cite{dinev2004internet} is a three-item Likert scale assessing concerns towards personal information provided on the Internet. The Identity Information Disclosure Scale \cite{stutzman2006evaluation} is specifically addressed to concerns regarding the disclosure of identity information in the the usage of social networks.

\subsection*{Aims and hypotheses of the present study}

The main objective of our study is to measure the relationship between the constructs of Sense of Community, Participation and Privacy compared to Culture and Sex. We examined the score on the scale for Sense of Community, Participation and Privacy in respect to Sex and Culture. We wanted to assess the effect of Culture and Sex on the perceived Sense of Community, Sense of Confidence/Concern toward the representation of the construct of Participation and Privacy (in order to consider cultural values referred to this construct, we provided an Italian/Turkish definition of the construct). In order to appreciate more how cultural differences are involved in representing the constructs of Participation and Privacy, we provided a semantic differential for both Italian and Turkish sub-samples. This measure has also been analyzed in respect to Sex. At the end of our analysis, we present the correlation structure between all the measures adopted for both the entire sample as well as the single sub-samples.

\subsection{Methods}

\subsection*{Participants}

A total of $180$ participants were recruited for the online questionnaires of our study. We divided the sample on the basis of their origin (i.e. Italy/Turkey).  
The Italian sub-sample ($29$ M; $54$F) shows a mean age of $24.69$ S.D. $3.9$ (Male= $24.24$ S.D. $4.46$; Female = $24.93$ S.D. $3.57$) and the Turkish sub-sample ($49$ M; $48$ F) shows a mean age of $22.4$ S.D. $2.67$ ( Male = $22.96$ S.D. $2.68$; Female = $21.83$ S.D. $2.55$). Most of the participants were students ($84.\%$) recruited from The University of Florence (Italian sub-sample) and the Cukurova University (Turkish sub-sample).

\subsection*{Procedures}

The main objective of this study, as mentioned above, is to measure the relationship between Sense of Community, Participation and Privacy, in two different cultures (Italian and Turkish). In order to assess this relationship in respect to Culture and Sex, a set of questionnaires was prepared.

The scale we used to asses Sense of Community was the union of two different scales: The Classroom and Community Inventory (CSCI) as found in Rovai, Wighting \& Lucking (2004), and the Sense of Community in School as found in Vieno, Santinello, Pastore \& Perkins (2007), for a total of $16$ items on a $5$ point Liker scale, from $1$= "strongly" disagree to $5$="Strongly agree" \cite{rovai2004classroom, vieno2007social}. Since this questionnaires are aimed to assess Sense of Community in a scholastic environment, we took in consideration only participants that at the moment of the administration were university students or university workers (e.g., teachers, researchers, PhD students). 

In order to obtain a measurement for both the Participation and Privacy constructs, two different questionnaires were developed. Both questionnaires were structured following this scheme: definition, semantic differential, and scale.

Definition: for both dimensions a general definition of the construct has been provided using those found in popular dictionaries  (Treccani dictionary for Italian, “Wikipedia” and “Türk Dil Kurumu “ for Turkish) for both Italian and Turkish sub-samples.

Semantic differential: the Semantic differential technique was used in order to appreciate more the cultural differences regarding both constructs \cite{osgood1964semantic}. According to the Osgood’s semantic differential (1964) a list of $10$ items made of two bipolar adjectives was provided (e.g., warm-cold; useful-useless; safe-dangerous). Participants were asked to choose where his/her position lie in the continuum (a Likert Scale from $1$ to $10$) from the positive to the negative side (adjective) of each of the $10$ items.

Scale: for both Participation and Privacy an ad-hoc questionnaire was developed. Both scales presented $10$ items assessing participants’ perception of Participation/Privacy in relation to different contexts (real life, online and social networks). Items were alternated into two sub-sections: $5$ items assessed the confidence towards these constructs, $5$ assessed concerns towards them.  Answers on a $5$-point Likert scale ($1$="Absolutely no", $2$="A few”, $3$="Moderately", $4$="A lot”, $5$="Very much”) were coded and summed up in order to obtain a general score for both Concerns/Confidence sub-scales.

\subsection*{Data analysis}

The data analysis activity was structured in two main phases. In the first one, we calculated the descriptive statistics, assessing the pre-conditions required by the subsequent inferential analysis. In particular, we checked the Gaussian distribution of the continuous variables, i.e. skewness and kurtosis $\in (-1;+1)$, and the sufficient balancing and size of the sub-samples of interest (i.e., sex and nationality). Then in the second phase we conducted the inferential analyses required by our main hypothesis.
In particular, within the context of this research the Pearson's \textit{r} correlation between the variables such as participation, privacy and sense of community were analyzed. Furthermore, an ANOVA analysis was conducted to find out if these three variables differentiate in terms of nationality and sex factors. SPSS.20 was used for these analyses.

\section*{Results}

In this section the main results of our research are presented. Descriptive statistics are presented first, where we show the results of the socio-demographic and operative variables (Participation, Privacy, Sense of Community) of our sample.  In the following section, we show results of the inferential analysis (i.e. correlation structures, ANOVA analysis, effects of socio-demographics on a semantic differential scale).

\subsection*{Descriptive statistics}

\subsubsection*{Sociodemographic variables}

The task has been administered to Italian and Turkish university students. The Italian sub-samples shows a mean age of $24.68$ and is composed by $54$ females and $29$ males, the Turkish people (mean age= $22.40$) by $48$ females and $49$ males.  

\begin{table}[h!]
\caption{Sociodemographic dimensions for the Italian and Turkish sub-samples. The standard deviation of each variable ($\sigma$) is reported between brackets}
\begin{center}
\small
\begin{tabular}{lccc}
\label{tab:descriptive_1}
Subsample&n&Age\\
\hline
Italian&83&24.68 (3.9)\\
Turkish&97&22.40 (2.7)\\
\hline
It Females &54&24.92 (3.6)\\
It Males &29&24.24 (4.5)\\
\hline
Tk Females &48&21.83 (2.6)\\
Tk Males &49&22.96 (2.7)\\
\hline
\end{tabular}

\normalsize
\end{center}
\end{table}

Most of the participants were students ($84\%$), single ($75\%$) and had an high degree of education (Bachelor degree= $71\%$). The Turkish sub-sample seem to be more uniform regarding civil status and education.

\begin{table}[h!]

\caption{Summary of descriptive statistics about the discrete sociodemographic variables of the sample}
\label{tab:Descriptive_0}
\begin{tabular}{lcccccc} \toprule
Variable & Total & \multicolumn{2}{l}{Italian} & \multicolumn{2}{l}{Turkish} \\
&& \multicolumn{2}{l}{Subsamples} & \multicolumn{2}{l}{Subsamples}\\ 
\cmidrule(r){3-6}
&& Female & Male & Female & Male \\ \midrule
\textbf{Occupation} & 180 &&&& \\
Student & 152 & 43&20&44&45 \\
Employed & 20 & 4&8&4&4 \\
Unemployed & 8 & 7&1&0&0 \\
\hline
\textbf{Civil status} & 180 &&&& \\
Single & 136 & 24&19&47&46 \\
Engaged & 38 & 29&6&1&2 \\
Married & 6 & 1&4&0&1 \\
\hline
\textbf{Education} & 180 &&&& \\
High Sc. & 24 & 12&11&0&1 \\
Bachelor & 128 & 27&11&44&46 \\
Mast. Sc & 28 & 15&7&4&2 \\
\hline\\
\end{tabular}

\end{table}

\subsubsection*{Operative variables}

Our scales about Participation and Privacy seem to be able to discern between the confidence and the concerns the participant has about these two dimensions (Table \ref{tab:Descriptive_2}). In the Italian sample, we found general lower values compared to the Turkish sub-sample, for both Confidence and Concerns. Average scores suggest that Italian females are more confident ($16.4$) than males ($15.4$) and less concerned (female=$12.9$; male= $12.1$) about participation. On the contrary they are more concerned ($14.9$) than males ($13.9$) and less confident (female=$12.1$; male=$13.3$) about privacy.
In the Turkish sample we see the same trend, but with higher values. Males and females have similar values about Participation Confidence (18.9 and 18.3), generally higher than the Concerns (female=$14.6$; males=$15.1$). Like in the Italian sample, Turkish participants have higher values in Privacy Concerns ($18.1$ for female and $16.9$ for male) than in Privacy Confidence ($13.6$ and $15.7$) and females show higher values than males on concerns for both measures.
Italian females ($55.17$) show higher values on the sense of community scale than Italian males ($51.38$). The Turkish sub-sample has generally lower values than the Italians, with similar scores for females ($50.23$) and males ($50.44$).

\begin{table*}[htbp!]
\centering
\captionsetup{justification=centering}
\small
\caption{Summary of descriptive statistics of the operative variables about participation and privacy}
\label{tab:Descriptive_2}
\begin{tabular}{lccccc} \toprule
Measure & \multicolumn{2}{l}{Italian} & \multicolumn{2}{l}{Turkish} \\
& \multicolumn{2}{l}{Subsamples} & \multicolumn{2}{l}{Subsamples}\\ 
\cmidrule(r){2-5}
& Female & Male & Female & Male \\ \midrule
\textbf{Participation} &&&& \\
Confidence & 16.4(3.3)&15.4(4.3)&18.3(3.9)&18.9(4.0) \\
Concerns & 12.9(3.6)&12.1(3.9)&14.6(2.6)&15.1(3.4) \\
\hline
\textbf{Privacy} &&&& \\
Confidence & 12.1(1.8)&13.3(3.0)&13.6(3.2)&15.7(3.0) \\
Concerns & 14.9(2.9)&13.9(2.8)&18.1(4.2)&16.9(4.5) \\
\hline
\textbf{Sense of } & &&&&\\
\textbf{Community} & 55.17(7.4)&51.38(12.3)&50.23(11.6)&50.44(11.1)\\
\hline
\end{tabular}
\end{table*}

\begin{figure}[h!]
\begin{center}
\includegraphics[width=0.45\textwidth]{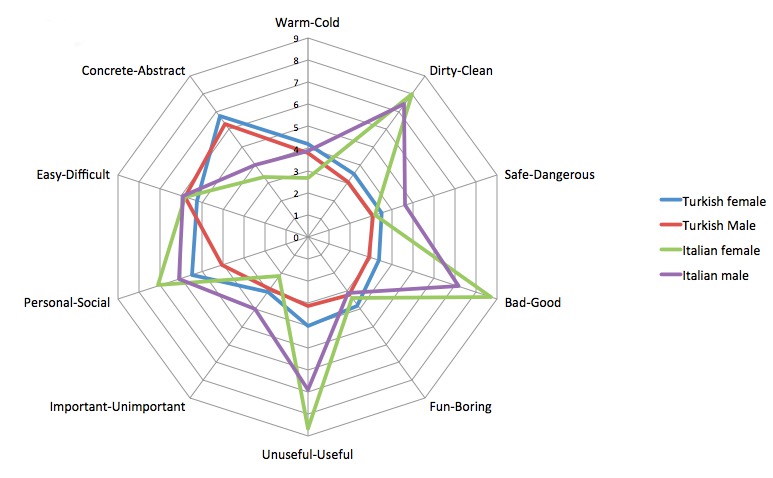}
\caption{\label{fig:Participation_differential} Semantic differential for all the sub-samples with respect to the concept of "Participation".}
\end{center}
\end{figure}

According to the Semantic differential, the most interesting results about Participation is the elevated contrast between Italians and Turkish People. 
As you can see in Figure \ref{fig:Participation_differential} Italians assessed almost all the adjectives at one of the extremes of the continuum; both males and females, showed very high scores for "Useless-Useful”, "Bad-Good”, "Dirty-Clean” and "Personal-Social”; for almost all other differentials the scores are very low, except for "Easy-Di cult”. Turkish samples present a totally different trend, where all values are low, only "Personal-Social”, "Easy-Di cult” and "Concrete-Abstract” have slightly higher values. For both Italians and Turkish people, the differences between males and females are minimal.

\begin{figure}[h!]
\begin{center}
\includegraphics[width=0.45\textwidth]{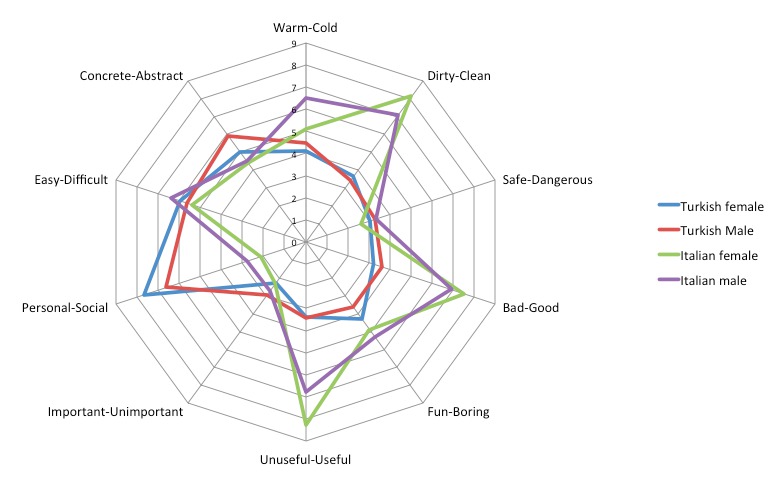}
\caption{\label{fig:Privacy_differential} Semantic differential for all the sub-samples with respect to the concept of "Privacy".}
\end{center}
\end{figure}

In the Privacy Semantic Differential \ref{fig:Privacy_differential} the Turkish sample shows a trend similar to the Participation one with mid-low values in almost all items, except for "Personal-Social”, "Easy-Difficult" and "Concrete-Abstract”. The Italian sample has some similarities, but we can see lower values for "Important-Unimportant”, whereas we found mid values in the "Easy-Difficult”, "Concrete-Abstract” and "Warm-Cold” items. Exactly like the Participation Semantic differential and even for privacy, the differences between males and females are very low in both samples. 

Finally, the sense of community scale reports higher values for Italians in general. In particular, females report higher values in the Italian sub-sample ($55.17$ vs. $51.38$), and lower but similar values in the Turkish sub-sample ($50.23$ vs. $50.44$), as indicated in Table \ref{tab:Descriptive_2}.

\subsection*{Inferential statistics}

The inferential analysis was structured in three different phases investigating the correlation structure characterizing the observable quantities taken into account in this study (i.e., age, education, sense of community, participation and privacy).  The second phase ran three ANOVA analyses in order to investigate the univariate and combined effects of nationality and sex on the three fundamental variables defining the constructs under scrutiny (i.e., sense of community, participation and privacy). Finally, the last phase was dedicated to assessing the effects of the socio-demographic factors of interest (i.e., nationality and sex) on the differential semantic dimensions describing the participation and privacy perception.\\

\subsubsection*{Univariate analysis: correlation between the order parameters}

\begin{table*}[htbp!]
\centering
\captionsetup{justification=centering}
\caption{Correlations between the dependent measures}
\begin{adjustbox}{angle=90}
\begin{tabular}{lccccccc} \toprule
\label{tab:Correlation_General}
Measure & Age & Education & Sense &\multicolumn{2}{l}{Participation} & \multicolumn{2}{l}{Privacy} \\
&&& of Community &\multicolumn{2}{l}{Submeasures} & \multicolumn{2}{l}{Submeasures} \\
\cmidrule(r){5-8}
&&&& Confidence & Concerns & Confidence & Concerns \\ \midrule
\textbf{Sense of Community} &&&& \\
It. Female & ns&$.28^*$&1&ns&ns&ns&ns \\
It. Male & ns&$-.45^*$&1&$.69^{**}$&$.54^{**}$&ns&$.58^{**}$ \\
Tk. Female & ns&ns&1&ns&ns&ns&ns \\
Tk. Male & ns&ns&1&$.39^{**}$&$-.33^*$&ns&ns \\
\hline
\textbf{Par. Confidence} &&&& \\
It. Female & ns&ns&ns&1&$-.53^{**}$&$.28^*$&ns \\
It. Male & ns&ns&$.69^{**}$&1&ns&ns&$.40^{*}$ \\
Tk. Female & ns&ns&ns&1&ns&$.61^{**}$&$.35^*$ \\
Tk. Male & ns&ns&$.39^{**}$&1&ns&$.40^{**}$&ns \\
\hline
\textbf{Par. Concerns} &&&& \\
It. Female & ns&ns&ns&$-.53^{**}$&1&ns&$.34^*$ \\
It. Male & ns&ns&$.54^{**}$&ns&1&ns&$.47^{**}$ \\
Tk. Female & ns&ns&ns&ns&1&ns&$.29^*$ \\
Tk. Male & ns&ns&$-.33^{*}$&ns&1&ns&$.40^{**}$ \\
\hline
\textbf{Pri. Confidence} &&&& \\
It. Female & ns&$-.31^{*}$&ns&$.28^{*}$&ns&1&ns \\
It. Male & ns&ns&ns&ns&ns&1&ns \\
Tk. Female & $.29^{*}$&ns&ns&$.61^{**}$&ns&1&ns \\
Tk. Male & ns&ns&ns&$.40^{**}$&ns&1&ns \\
\hline
\textbf{Pri. Concerns} &&&& \\
It. Female & ns&$-.40^{**}$&ns&ns&$.34^{*}$&ns&1 \\
It. Male & ns&ns&$.58^{**}$&$.40^{*}$&$.47^{**}$&ns&1 \\
Tk. Female & ns&ns&ns&$.35^{*}$&$.29^{*}$&ns&1 \\
Tk. Male & ns&ns&ns&ns&$.40^{**}$&ns&1 \\
\hline
\multicolumn{8}{l}{Pearson \textit{r} correlation, *: \textit{p} < .05, **: \textit{p} < .01, \textit{ns}: not significant.}
\end{tabular}
\end{adjustbox}
\end{table*}

The Pearson's \textit{r.} correlation statistic was adopted to estimate the relations between the fundamental variables of interest with respect to the $4$ sub population considered by our study \ref{tab:Correlation_General}. 

The age of the participants appears to have no significant effects. The only exception is related to the Turkish female sample where a weak relation emerged between age and privacy confidence.

Education appears to be related to the sense of community only in the Italian sample. In particular, it showed quite an interesting reversed effect regarding sex for Italian males, where we obtained a negative correlation explaining about $20\%$ of the variance which suggested that males with a higher education are characterized by a lower sense of community. However, the Italian females reported a positive correlation between the two measures explaining the $8\%$ of the variance. For what concerns the privacy construct, education appears to affect only the Italian female perception. In particular, we always obtained a negative correlation accounting for $9\%$ and $16\%$ of the variance respectively for privacy confidence and privacy concerns.

The sense of community correlation structure shows interesting features as well. First of all no significant correlation appears to relate the sense of community with privacy and participation for the females disregarding nationality. On the contrary, the male sub-samples both reports positive correlations between sense of community and participation confidence with higher values for Italians ($\textit{r.} = 0.69$), and lower values for Turkish students($\textit{r.} = 0.39$). While an opposite behavior appears to characterize the relation with the participation concerns, with values respectively of $\textit{r.}=0.54$ for Italians, and of $\textit{r.}=-0.33$ for Turkish males. Only the Italian males show a significant positive correlation between sense of community and privacy concerns ($\textit{r.}=0.58$). In order to evaluate the non-trivial relationships existing between concerns and confidence for both the participation and privacy perceptions, we explicitly took into account all the correlation structures describing such sub-measures separately.

The participation confidence appears to be significantly and negatively related to the participation concerns only within the Italian females sub-sample, explaining $28\%$ of the variance. For what concerns the relationship between participation and privacy confidence, all the sub-samples report positive correlations with the only exception of the Italian males, suggesting how an increment in the first produces a coherent change in the second. In particular, the Turkish females show the higher correlation ($\textit{r.}=0.61$), and the Turkish males the second one ($\textit{r.}=0.40$), with the Italian females that report the weakest relation ($\textit{r.}=0.28$). Finally the participation confidence is even related with the privacy concerns for Italian males ($\textit{r.}=0.40$), and Turkish females ($\textit{r.}=0.35$).
In brief, the participation concerns are reported to correlate with the privacy concerns, and not with the privacy confidence for all the investigated sub-samples. All the correlations here are between $11\%$ and the $22\%$.\\

\subsubsection*{Sense of community: nationality and sex effects}

The effects of nationality and sex, as well as their interaction on the sense of community were evaluated by means of ANOVA analysis. As reported in table \ref{tab:Senseofcommunity_1}, no differences emerge regarding the sex, nor the interaction between nationality and sex. The only factor affecting such a dimension appears to be the nationality, even if the effect is very moderate explaining just $3.5\%$ of the variance, with the Italian sub-sample reporting a higher value.

\begin{table}[h!]
\centering
\caption{In table the best ANOVA model for the dependent variable Sense of Community, considering the factors sex and nationality, is reported.}
\label{tab:Senseofcommunity_1}
\begin{tabular}{lccc} \toprule
\multicolumn{3}{l}{\textbf{Sense of community}} \\
Factor & \textbf{F} &\textit{Sign}& $\eta^2$\\
\hline
Nationality & 6.4 & \textit{p.} < 0.05 & 0.035 (3.5 \%)\\
Sex & 1.2 & \textit{ns} & 0.007 (0.7\%)\\
Nat*Gen & 1.5 & \textit{ns} & 0.009 (0.9\%)\\
\hline
\hline

\end{tabular}
\end{table}

\subsubsection*{Participation dimensions: nationality and sex effects}

Confidence and concerns sub-measures concerning participation were separately analyzed by ANOVA (table \ref{tab:Participation_1}), in order to estimate the role of nationality and sex on the two dimensions. For both categories, sex and interaction vs sex and nationality do not have statistically significant effects, while again nationality appears to be the only factor impacting on such dimensions. In both the cases, the cultural differences seem to account for approximately $10\%$ of the variance. In particular, both the participation confidence as well as the participation concerns appear to be significantly higher within the Turkish sample.

\begin{table}[h!]
\centering
\caption{In table the best ANOVA model for the variables connected to the participation dimension (i.e., confidence \& concerns) are presented, considering the factors sex and nationality}
\label{tab:Participation_1}
\begin{tabular}{lccc} \toprule
\multicolumn{3}{l}{\textbf{Participation Confidence}} \\
Factor & \textbf{F} &\textit{Sign}& $\eta^2$\\
\hline
Nationality & 20.7 & \textit{p.} < 0.01 & 0.105 (10.5 \%)\\
Sex & 0.1 & ns & 0.001 (0.1\%)\\
Nat*Gen & 1.8 & ns &  0.010 (1\%)\\
\hline
\multicolumn{3}{l}{\textbf{Participation Concerns}} \\
Factor & \textbf{F}& $\eta^2$\\
\hline
Nationality & 20.1 & \textit{p.} < 0.01 & 0.102 (10.2 \%)\\
Sex & 0.1 & ns & 0.001 (0.1\%)\\
Nat*Gen & 0.6 & ns & 0.009 (0.9\%)\\
\hline
\hline

\end{tabular}
\end{table}

\subsubsection*{Privacy dimension: nationality and sex effects}

In table \ref{tab:Privacy_1} the best predicting model linking the nationality and sex factors to the sub-scales related to the privacy dimension (i.e., confidence and concerns) is reported. In these cases, we obtained a significant effect both on nationality as well as on sex, and again no effects were revealed about the interaction between the factors of interest. In particular, nationality explains $11.4\%$ for the privacy confidence, and the $14.1\%$ of the privacy concerns. In both cases, again the Turkish sample is characterized by higher values. In regards to sex, females reported higher values for privacy concerns, and lower values for privacy confidence, respectively explaining $8.3\%$ and $2.1\%$ of the total variance.

\begin{table}[h!]
\centering
\caption{In table the best ANOVA model for the variables connected to the privacy dimension (i.e., confidence \& concerns) are presented, considering the factors sex and nationality}
\label{tab:Privacy_1}
\begin{tabular}{lccc} \toprule
\multicolumn{3}{l}{\textbf{Privacy Confidence}} \\
Factor & \textbf{F} &\textit{Sign}& $\eta^2$\\
\hline
Nationality & 22.6 & \textit{p.} < 0.01  & 0.114 (11.4 \%)\\
Sex & 15.8 & \textit{p.} < 0.05 & 0.083 (8.3\%)\\
Nat*Gen & 1.3 &  ns & 0.007 (0.7\%)\\
\hline
\multicolumn{3}{l}{\textbf{Privacy Concerns}} \\
Factor & \textbf{F}& $\eta^2$\\
\hline
Nationality & 28.9 & \textit{p.} < 0.01 & 0.141 (14.1 \%)\\
Sex & 3.7 & \textit{p.} < 0.05 & 0.021 (2.1\%)\\
Nat*Gen & 0.1 & ns & 0.001 (0.1\%)\\
\hline
\hline

\end{tabular}
\end{table}

\subsubsection*{Semantic differential about Participation and Privacy}

Very interesting results emerged from the semantic differential analysis regarding the nationality and sex as well as  their combined effect. Table \ref{tab:Semantic_Differentials_models} reports only the significant effects, that nevertheless appear to be present in both cases (i.e., semantic differential related to privacy and participation separately) for $8$ dimensions out of the original $10$. Some dimensions are characterized by moderate effects, with an explained variance ranging between $2.6\%$ and $9.6\%$. It is worth noting some dimensions appear as strongly affected by the nationality factor. For instance the concept of participation appears to elicit impressive differences, with "bad-good" accounting for $61.2\%$ of the total variance, "dirty-clean" for $50.2\%$, and "unusefull-usefull" for $46.5\%$. Even the privacy concept appears differently represented by the two cultural samples, with "personal-social" dimension explaining $45.8\%$, "unuseful-useful" $45.2\%$, "dirty-clean" $44.1\%$, and "bad-good" $40.4\%$. In particular, as reported in figures \ref{fig:Participation_differential} and \ref{fig:Privacy_differential}, the Italian sample seems to perceive the participation concept as more "clean", "good" and "useful" compared to the Turkish sample. The same effects are reported for the privacy concept with the only exception of the dimension "personal-social", in which the Turkish sample perceives "privacy" as more social than the Italians.

\begin{table}[h!]
\centering
\small
\caption{In table the best ANOVA models for the variables connected to the semantic differentials about both Participation and Privacy are presented, considering the factors sex and nationality. Statistical significance (\textit{p.}) is reported in table  as follows: $^*: <0.05$ and $^**:< 0.01$}
\label{tab:Semantic_Differentials_models}
\begin{tabular}{lccc} \toprule
\multicolumn{4}{l}{\textbf{Participation}} \\
Dimension & Factor & \textbf{F} & $\eta^2$\\
\hline
Warm-Cold & Nationality & 4.7* & 0.026 (2.6 \%)\\
& Nat*Gen & 5.5* &  0.031 (3.1\%)\\
\hline
Dirty-Clean & Nationality & 177.6** & 0.502 (50.2 \%)\\
\hline
Safe-Dangerous & Nat*Gen & 7.3** &  0.040 (4.0\%)\\
\hline
Bad-Good & Nationality & 277.8** & 0.612 (61.2 \%)\\
& Sex & 12.3** & 0.065 (6.5\%)\\
\hline
Unuseful-Useful & Nationality & 153.0** & 0.465 (46.5 \%)\\
& Sex & 15.2* & 0.080 (8.0\%)\\
\hline
Imp.-Unimportant & Nationality & 7.4** & 0.041 (4.1 \%)\\
& Nat*Gen & 11.2** &  0.060 (6.0\%)\\
\hline
Personal-Social & Nationality & 18.8** & 0.096 (9.6 \%)\\
& Sex & 8.4** & 0.045 (4.5\%)\\
\hline
Concrete-Abstract & Nationality & 63.9** & 0.266 (26.6 \%)\\
\hline
\hline
\multicolumn{3}{l}{\textbf{Privacy}} \\
Dimension & Factor & \textbf{F} & $\eta^2$\\
\hline
Warm-Cold & Nationality & 12.6** & 0.067 (6.7 \%)\\
& Sex & 4.5* &  0.025 (2.5\%)\\
\hline
Dirty-Clean & Nationality & 138.6** & 0.441 (44.1 \%)\\
\hline
Bad-Good & Nationality & 199.4** & 0.404 (40.4 \%)\\
\hline
Fun-Boring & Nationality & 9.1** & 0.049 (4.9 \%)\\
\hline
Unuseful-Useful & Nationality & 145.2** & 0.452 (45.2 \%)\\
& Sex & 4.2* & 0.023 (2.3\%)\\
& Nat*Gen & 5.2* &  0.028 (2.8\%)\\
\hline
Imp.-Unimportant & Sex & 4.1* & 0.023 (2.3 \%)\\
\hline
Personal-Social & Nationality & 148.6** & 0.458 (45.8 \%)\\
& Nat*Gen & 4.9* & 0.027 (2.7\%)\\
\hline
Concrete-Abstract & Nationality & 5.5* & 0.030 (3.0 \%)\\
\hline
\hline

\end{tabular}
\end{table}

\section*{Discussion}

The psychological constructs of Participation and Privacy are currently among the most interesting psychological concepts to investigate. The growing multi-disciplinary literature about them, ranging from engineering to complex systems and social sciences, witnesses the impact of ICT on modern human society, and in general on every new kind of interactions among people and groups \cite{castellano2009statistical,Vilone201684,cecconi2016new}. Many societal dynamics, from the microscopic interaction within families to the macroscopic cultural and sociological movements involving people from all over the world, are based on the same human tendencies of interaction with others by means of the adoption of cultural and personal dependent strategies of coping. From the "Arab spring" in 2010, to the "Occupy Wall Street movements" in 2011, and the recent protests in Turkey (i.e., from Taksim square protests to the recent coup), internet has been considered by many commentators as one of the fundamental ingredients. Of course the connection between people, especially within the younger generations, has provided new incentives and mainstreams that have nurtured the protest movements. On the other hand, such events could be considered as a direct consequence of the changes produced by internet on its users. Actually, from this point of view it seems very hard to clarify the role of the world wide web on such cultural and societal changes, while it seems more reasonable to take into account both the dynamics described above. Human beings change in order to satisfy environmental requirements (i.e., increasing their adaptation), to negotiate social constructs, norms and behaviors useful to communicate, interact and solve problems with others as well as represent elements affecting and actively changing the environment they belong to.
 For instance, the spreading of social networks in the last few years required non-digital natives to make a great effort to adapt to such a brand new social order, while it profoundly shaped the social skills and attitudes of digital natives. Such a double process (i.e., to adapt old strategies to new tasks, and to adopt new concepts and habits) is not just a question of age (i.e., digital natives vs non-digital natives), but it represents a double process acting in different degrees within every participant, and in different ways (i.e., with different speed and effectiveness) along the different cultures, and socio-demographic clusters. In general, it is expected that the dynamics of such changes will define new interesting scenarios and phenomena, and of course both the spreading of new concepts and social negotiations of old habits and norms will continue to determine societal as well as personal changes.

The hypotheses of our study moved from the basic assumption that the "fluid" human perception of socially-negotiated concepts like Participation and Privacy, can differ a lot between two different cultural groups, depending on structural differences of the cultures and on different recent events affecting the university students participating in the survey.

The Turkish and the Italian cultures have ancient common roots, and could be both considered as European cultures even if with a very different recent history. Finally, the different role of sex within the two cultural groups was considered and evaluated in the multivariate analysis.

The two sub-samples appeared as comparable regarding the size, average age and sex balance. With the age that actually presented very little dispersion around the average values, and that can be considered as a constant describing a very specific social cluster (i.e., those of the university students from a medium size city). Consequently, it is not surprising that no significant correlations emerged concerning age. The only correlation with the age variable is reported in the Turkish female sub-sample as affecting the Privacy Confidence. The older the participant is the greater is her confidence and trust in the concept of Privacy (Tab. \ref{tab:Correlation_General}).

The education variable represents the number of formal years of study accomplished by the participants, and the different distribution of the two sub-samples is due only to the different organization of the two educational systems. For this reason only the Italian sample presents significant correlations between the education variable and the operative dimensions. The opposite relationship between sex and sense of community is very interesting. In the Italian male sub-sample a negative correlation is revealed, while in the female one a significant positive correlation is present. Moreover in the Italian sample, the males are characterized by a negative correlation between sense of community and education, while the Italian females show the opposite trend. The Privacy sub-dimensions, confidence and concerns, show a negative correlation in age only for Italian females, and consequently appear to reduce both their confidence and concerns about Privacy with the increase of education.

The sense of community dimension shows  very interesting relation patterns within our sample, reporting significant correlations only for the male sub-samples, and not always in the same direction. The Participation concept is perceived as more positive and trustworthy as the sense of community increases, the same effect acts on Italian male sub-samples even for the concerns, while it presents an opposite effect on the Turkish male sub-sample (i.e., the concerns about participation decrease with the increase of sense of community). Finally the Italian sub-sample presents a positive correlation between the sense of community and Privacy concerns scores (Tab \ref{tab:Correlation_General}).

In general, it is interesting to observe how the results of the correlation matrix  between Participation and Privacy measurements are quite empty. In particular, larger correlations were expected between the confidence and concerns sub-measures. Such  data indicates the multi-structured and complex architecture of the concepts under scrutiny and how in different cultures their interplay could not be easy to reveal or predict.

In order to evaluate the interplay between sex and culture on the operative variables of our study (i.e., sense of community, participation and privacy concerns and confidence), we conducted a series of separate ANOVA in order to test the effects of the factors as well as of their interaction.

The sense of community does not appear to differ much regarding the nationality ($3,5\%$), and the sex does not play a significant role in the interaction between sex and nationality.  In other words, the two cultures appear to be comparable in terms of sense of community.

On the contrary, more relevant differences are detected regarding the Participation and Privacy measures. The ANOVA investigating the sex and nationality effects on the sub-scores of the Participation scale (Tab. \ref{tab:Participation_1}) reports that around the $10\%$ of the variance is explained by the nationality factor. In particular, both the confidence and the concerns about participation are significantly higher in the Turkish sample. A greater portion of the variance is explained by the ANOVA investigating such effects on the Privacy sub-scores. Concerning this construct respectively $11,4\%$ and the $14,1\%$ are associated with the nationality, again indicating the Turkish sub-sample as more confident and concerned about Privacy (Tab. \ref{tab:Privacy_1}). Finally an effect of the sex is revealed as explaining respectively the $8,3\%$ and the $2,1\%$ of the variance of Privacy confidence and concerns sub-scores, with the Females reporting always a lower score. 

The most fascinating results are those produced by the semantic differential measures. Even in this case a series of ANOVA were conducted analyzing every dimension presented by the semantic differential, considering the sex and the nationality as independent factors of variance. The Fisher’s Fs and the Etas squared reported in table 8 delineates impressive differences in the perception of such concepts always mainly related to the nationality (i.e., the culture).

Both the perception of the Participation and Privacy appear to be affected by nationality and sex; in particular in 8 cases out of 16 the explained variance (i.e., the entity of the difference explained by the factors) is under $10\%$. Surprisingly the other 8 cases report an explained variance ranging between the $61,2\%$ and the  $26\%$. 

For what concerns Participation, it seems that Italians perceive it as a more "clean", "good", "useful" and "social" than Turkish, with no significant difference to sex. Among the samples, there are no significant differences on the perceived importance, warmness, easiness and  perceived "fun" related to the experience of participating. The fact that Turkish people perceive it as more "dirty", "concrete", "bad", "useless" could be related to the recent experiences of the Taksim square protests in which really violent clashes happened.

As regards to the concept of Privacy, it seems that the Italians perceive it as a more "useful", "good", "clean" and "personal" concept than the Turkish students. In order to appreciate these results with a cultural interpretation, we wanted to highlight the key role that the recent Italian mass media and political debates have given to the concept of privacy. In fact, the Italian political agenda in the recent years has been highly focused on a privacy regulation, especially regarding important legal topics. This concern has led to a more developed concept of privacy, highly related to the concept of "human rights that need to be respected and protected" as found in the definition provided by the main Italian dictionary (Treccani dictionary). This concept of privacy as a right is not stressed at the same level for example in the current English definition, which stresses more the perspective of an "ability of an individual or group to seclude themselves" (Wikipedia), demonstrating how this concept is still in development and culturally shaped and variable. The fact that the Turkish students perceive the construct of Privacy as more "useless", "social", "dirty" and "bad" could be given to the fact that this concept is quite new to them. In fact, the word "privacy" is not in the main Turkish dictionaries at the moment and even the literal translation of this term in this language ("gizlilik") needed a well-structured explanation to make it comprehensible to the Turkish sub-sample.
We believe that a better understanding of the dynamics underneath the use of ICT tools can provide us the necessary knowledge in order to shape these environments in a more efficient way (i.e. taking into account cultural, sex and historical differences), strengthening positive features and mitigating the negative ones. 

\section*{Acknowledgments}

This work was funded by the European Commission under the FP7-ICT-2013-10 call, proposal No. 611299, project SciCafe 2.0, and by H2020 FETPROACT-GSS CIMPLEX Grant No. 641191.

\bibliographystyle{apacite}
\bibliography{BIBLIO}

\end{document}